# Tunable resins with PDMS-like elastic modulus for stereolithographic 3D-printing of multimaterial microfluidic actuators†

Alireza Ahmadianyazdi, [iD] *[a] Isaac J. Miller[b] and Albert Folch[a]

Stereolithographic 3D-printing (SLA) permits facile fabrication of high-precision microfluidic and lab-on-a-chip devices. SLA photopolymers often yield parts with low mechanical compliancy in sharp contrast to elastomers such as poly(dimethyl siloxane) (PDMS). On the other hand, SLA-printable elastomers with soft mechanical properties do not fulfill the distinct requirements for a highly manufacturable resin in microfluidics (e.g., high-resolution printability, transparency, low-viscosity). These limitations restrict our ability to print microfluidic actuators containing dynamic, movable elements. Here we introduce low-viscous photopolymers based on a tunable blend of the monomers poly(ethylene glycol) diacrylate (PEGDA, $M_w$ ~ 258) and the monoacrylate poly(ethylene glycol) methyl ether) methacrylate (PEGMEMA, $M_w$ ~ 300). In these blends, which we term PEGDA-co-PEGMEMA, tuning the PEGMEMA content from 0% to 40% (v/v) alters the elastic modulus of the printed plastics by ~400-fold, reaching that of PDMS. Through the addition of PEGMEMA, moreover, PEGDA-co-PEGMEMA retains desirable properties of highly manufacturable PEGDA such as low viscosity, solvent compatibility, cytocompatibility and low drug absorptivity. With PEGDA-co-PEGMEMA, we SLA-printed drastically enhanced fluidic actuators including microvalves, micropumps, and microregulators with a hybrid structure containing a flexible PEGDA-co-PEGMEMA membrane within a rigid PEGDA housing. These components were built using a custom "Print-Pause-Print" protocol, referred to as "3P-printing", that allows for fabricating high-resolution multimaterial parts with a desktop SLA printer without the need for post-assembly. SLA-printing of multimaterial microfluidic actuators addresses the unmet need of high-performance on-chip controls in 3D-printed microfluidic and lab-on-a-chip devices.

## 1. Introduction

Microfluidic actuators are essential components for the integrated control and transport of liquids within microfluidic and lab-on-a-chip devices.[1–4] Microactuators such as microvalves have been used for multiplexing,[1] sorting/separating,[2] pumping,[4] and mixing[5,6] of fluids in microchannel networks. Microfluidic valves have also been operated analogously to microelectronic components (e.g., transistors) to enable automated liquid manipulation or on-chip processing such as logic operations,[7] signal amplification,[8] and frequency-dependent response,[9] among others.

The function and performance of fluidic actuators critically relies on the elastic behavior of their dynamic, movable elements such as a membrane or diaphragm.[1,10] Thus, in conventional microfabricated actuators, poly(dimethyl siloxane) (PDMS) has been the primary choice among microfluidic engineers due to its advantageous elasticity and softness,[1,8] among other favorable properties: the material is inexpensive, optically clear, and biocompatible; its molding procedure is safe and easy to learn[11] and can be used to recreate biomimetic tissue-like properties, as with lung-on-a-chip.[12] However, the large-scale dissemination (including commercialization and non-profit distribution) of PDMS-molded microdevices remains challenging because PDMS molding is a labor-intensive procedure. In addition, PDMS is porous and hydrophobic, so both absorption into PDMS[13,14] and adsorption onto PDMS can potentially alter experimental outcomes by changing the target concentrations and by partitioning molecules in undesired regions of a microfluidic device.[15] Although coatings such as sol–gel[16] or silicate glass[17] can mitigate this problem, they add substantial processing time/complexity to the prototypes and modify the elasticity of PDMS.

[a] Department of Bioengineering, University of Washington, Seattle, WA, 98105, USA. E-mail: aahmad33@uw.edu
[b] Department of Chemical Engineering, University of Washington, Seattle, WA, 98105, USA

† Electronic supplementary information (ESI) available: Mechanical behavior of PEGDA-co-PEGMEMA plastics; pressure-displacement data for a 40% PEGDA-co-PEGMEMA membrane; rheological behavior of PEGDA and PEGMEMA monomers; creep and fatigue of PEGDA-co-PEGMEMA plastics; contact angle measurements; print-pause-print method; PEGDA-co-PEGMEMA curing behavior; experimental setup for the evaluation of microvalves; high-resolution microvalves; micropump operation and evaluation setup; STL design files. See DOI: https://doi.org/10.1039/d3lc00529a

High-precision stereolithographic (SLA) 3D-printing of microfluidic devices promises to alleviate some of the above limitations of PDMS fabrication. By enabling high-resolution, semi-automated fabrication from a digital file, SLA-printing shortens the time from prototype to product, lowers manufacturing and distribution costs, and enables the customization of complex-architecture microfluidic devices.[18–21] In recent years, several groups including ours have SLA-printed all-plastic, transparent, high-resolution, durable, and cytocompatible microfluidic chips using photopolymer resins based on low molecular weight PEGDA ($M_w \sim 258$)[22–30] (all references to "PEGDA" below, unless specified otherwise, refer to the 3D-printed version of PEGDA with $M_w \sim 258$). However, the popularity, reliability, and mechanical performance of plastic PEGDA has not yet achieved that of PDMS, hindering its practical use by a broader audience of researchers. The Young's modulus of photopolymerized PEGDA is around 870 MPa, about two orders of magnitude greater than that of PDMS,[31] and the maximum elastic strain (= $\Delta L/L_0$ with $L_0$ being the initial unstretched length and $\Delta L$ the elastic elongation) of PDMS is several times larger than that of PEGDA.[32] These characteristics limit the application of PEGDA for the efficient design of microfluidic actuators in which the flexuring element must be made from a highly deforming, elastic material in analogy to soft-lithographic actuators based on PDMS.[1] On the other hand, SLA-printable elastomers with similar mechanical properties to PDMS[33–38] (soft, highly stretchable), are several times more viscous than PEGDA (due to high-$M_w$ monomers and oligomers), have limited printing resolution, or are optically opaque, restricting their use to SLA print transparent, and high-resolution voids and microchannels needed in microfluidics. Thus, there is a need for photopolymers with softer mechanical properties than PEGDA, yet similar printability, cytocompatibility, solvent compatibility, and low drug absorptivity necessary for many microfluidic applications (*e.g.*, drug-screening, cell-based assays).

As PEGDA in microfluidics is usually formed by photopolymerization, we sought to find a monomer that, added to the PEGDA resin, would co-polymerize with PEGDA and modulate its elastic behavior while maintaining its other desirable properties. Beamish *et al.*[39] modified the mechanical properties of high molecular weight PEGDA-based hydrogels by co-polymerizing the PEGDA monomers ($M_w \sim 6000$) with mono-acrylate ones. They found that addition of PEGMA (PEG-mono-acrylate) monomers ($M_w \sim 2000$ or 5000) to PEGDA ($M_w \sim 6000$) enhanced the shear modulus and the cross-linking of the resulting hydrogel network.[39] They attributed these property changes to the increased kinetic chain length of poly(acrylic acid) nodes, unreacted acrylate moieties, and entanglement in the cross-linked network solely by the presence of PEGMA. Based on Beamish *et al.*'s findings, we hypothesized that co-polymerization of low-$M_w$ PEGDA in the presence of a mono-methacrylate monomer with similar $M_w$ would modify the mechanical properties of PEG-derived plastics. The similarity in $M_w$ allows us to exclude a contribution of high $M_w$ chains in the mechanical strength of the polymer, as explained in ref. 40. As a confirmation of our hypothesis, here we describe the use of PEGMEMA (PEG-methyl ether methacrylate) monomers at $M_w \sim 300$, a mono-methacrylate herein referred to as PEGMEMA, that upon co-photopolymerization with PEGDA monomers at various mixing ratios can extensively tune the elastic modulus of PEGDA plastics by over $\sim 2$ orders of magnitude. As a result, we were able to yield plastics with a Young's modulus as low as $\sim 2.2$ MPa comparable to that of PDMS,[41] and $\sim 400$ times smaller than that of plastic PEGDA. We refer to this newly introduced photopolymers as PEGDA-*co*-PEGMEMA family, indicating that a variety of resins can be prepared by tuning the mixing ratio between PEGMEMA and PEGDA monomers. PEGDA-*co*-PEGMEMA blends are highly manufacturable resins for microfluidics, since they have similar physical characteristics to the PEGDA resins widely used to 3D-print microfluidic devices.[22,24,25] This similarity stems from the comparable density (PEGDA with $M_w \sim 258$: 1.11 mg mL$^{-1}$; PEGMEMA with $M_w \sim 300$: 1.05 mg mL$^{-1}$; obtained from the manufacturer, see Experimental section), low viscosity (PEGDA with $M_w \sim 258$: $\sim 13.86$ cp; PEGMEMA with $M_w \sim 300$: $\sim 10.88$ cp; see Fig. S3†), and transparency of PEGDA and PEGMEMA resins alone. Importantly, since both monomers contain either acrylates or methacrylates, any selected combination of PEGDA-*co*-PEGMEMA resins can bond strongly together by photopolymerization. Accordingly, this family of resins can be used to fabricate monolithic, multimaterial parts by multimaterial SLA-printing methods[42,43] including a modified version of our "Print-Pause-Print" protocol,[44,45] here termed as "3P-printing". 3P-printing is compatible with commercially available SLA machines without the need for custom-built process control. Hence, using a desktop DLP-based (digital light projector) SLA-printer, we 3P-printed the tunable PEGDA-*co*-PEGMEMA resins into high-performance fluidic actuators with a hybrid architecture comprising of flexible PEGDA-*co*-PEGMEMA membranes within a rigid PEGDA structure. These actuators include microvalves, peristaltic micropumps, and passive microregulators with superior performance to their conventionally 3D-printed single-material counterparts.

## 2. Results and discussion

### 2.1. Characteristics of PEGDA-*co*-PEGMEMA

Here we introduce PEGDA-*co*-PEGMEMA as a family of acrylate-based photopolymers for SLA-printing of flexible plastics. PEGDA-*co*-PEGMEMA resins are prepared from a tunable mixture of PEGDA and PEGMEMA monomers, a photo-initiator, and a photo-absorber (see Experimental section). Fig. 1 validates our hypothesis that addition of a low-$M_w$ mono-methacrylate PEGMEMA can alter the stiffness of PEGDA plastics, and yield highly flexible polymers. In Fig. 1a, we illustrated the radical photopolymerization scheme of PEGDA (top), and PEGDA-co-PEGMEMA (bottom). In the absence of PEGMEMA, photopolymerization of PEGDA in the presence of a photo-initiator creates a densely cross-linked network of short polymer chains leading to a dry, impermeable plastic at least 400 times stiffer than PDMS. When PEGMEMA is added (bottom), the acrylate-free end of this monomer (shown by circles in Fig. 1a) fails to cross-link during radical polymerization, resulting in smaller degree of cross-linking and a looser network with dangling polymer chains that are susceptible to slide.

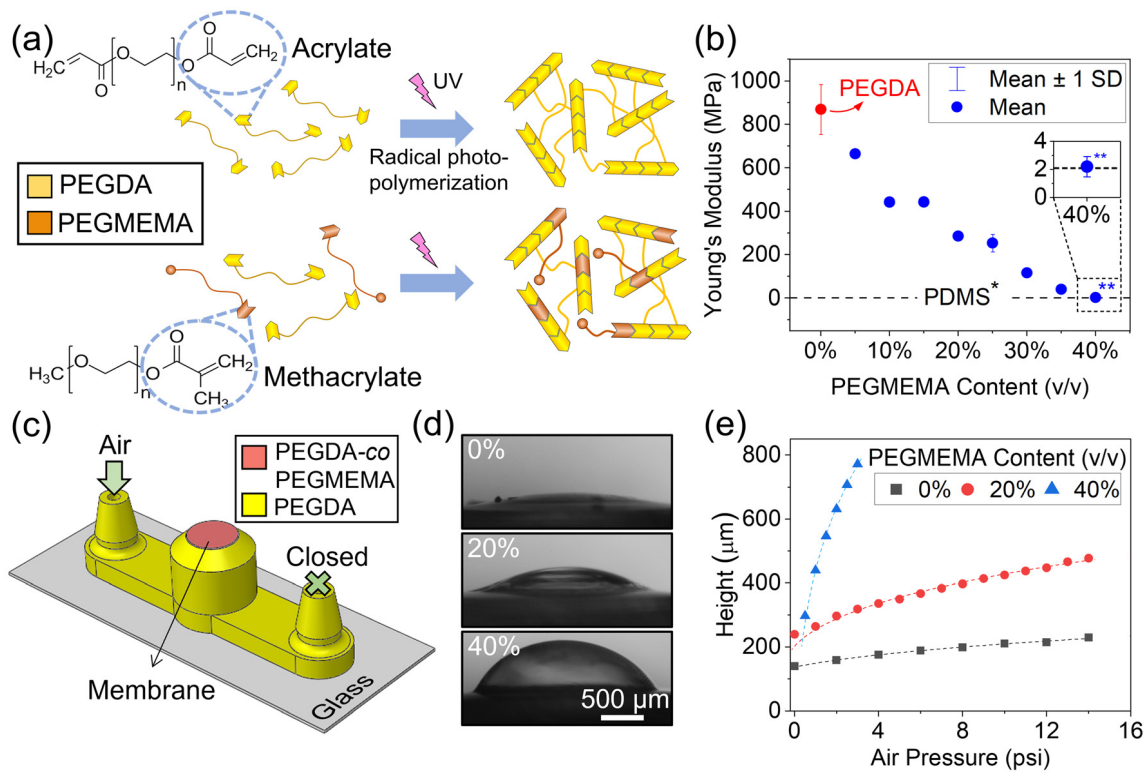

Fig. 1 Characterization of PEGDA-co-PEGMEMA resins. (a) Radical photopolymerization schemes of PEGDA (top), and PEGDA-co-PEGMEMA (bottom). (b) Young's modulus of PEGDA-co-PEGMEMA derived plastics (PEGDA at $M_w \sim 258$, and PEGMEMA at $M_w \sim 300$). *PDMS Young's modulus presented as a dashed horizontal line at ∼2 MPa.[31] The inset shows a magnified view of the 40% PEGDA-co-PEGMEMA data point. **Estimated from pressure-displacement data in Fig. S2.† (c) 3D CAD model of a pressurization chamber 3P-printed in PEGDA (yellow) to measure the deflection of a PEGDA-co-PEGMEMA membrane (red), illustrating the pneumatic inlet (green arrow) and the blocked outlet (green cross). (d) Micrographs of pressurized membranes made from 0%, 20%, and 40% PEGDA-co-PEGMEMA at 2 psi actuation. The membranes have an average thickness of 60 ± 10 μm and a diam. of 2 mm. (e) Graph depicting the maximum deflection height versus actuation pressure for the three membranes shown in image (d). The dashed lines present the fitted curves ($R_{avg}^2 \sim 0.98$) according to elastic membrane deflection in eqn (S1).†[48] Note that the deflection at zero air pressure is non-zero due to initial slack in the membranes caused by the SLA printing process.

To assess the effect of mono-methacrylate concentration on polymer elasticity, we measured the Young's moduli of PEGDA-co-PEGMEMA parts prepared with different ratios of PEGMEMA : PEGDA monomers. Hereafter, the volumetric content (v/v) of PEGMEMA in PEGMEMA : PEGDA mixture appears as a percentage before "PEGMEMA" for the resin blends, and as a percentage before "PEGDA-co-PEGMEMA" for the polymerized resin. We first prepared a set of PEGDA-co-PEGMEMA resins from a monomer precursor containing 0 to 40% PEGMEMA in PEGDA in increments of 5%. With these photopolymers, we SLA-printed dog-bone specimens (Fig. S1a†) and measured their elastic behavior under in-plane tensile load (see Experimental section). Fig. 1b plots the Young's moduli of these specimens with respect to the amount of PEGMEMA in their constituent resins. As shown, increasing PEGMEMA in PEGDA-co-PEGMEMA photopolymers from 0 to 35% decreased the Young's modulus of the plastic by a factor of ∼22, achieving values around 40 MPa for 35% PEGMEMA. A further increase in the PEGMEMA content from 35% to 40% yielded plastics with a Young's modulus of ∼2.2 MPa, similar to PDMS (as reported in ref. 31 and 41). Note that, as explained in the Experimental section, the Young's modulus of 40% PEGDA-co-PEGMEMA was approximated using the analysis of membrane deflection presented in Fig. S2.† These results suggest that, analogously to what occurs in high-$M_w$ hydrogels,[39] the presence of PEGMEMA chains in the PEGDA-co-PEGMEMA polymer loosens the otherwise packed PEGDA plastic network, allowing for sliding of the unbound PEGMEMA ends which leads to a more compliant and flexible polymer.

The contribution of $M_w$ to elasticity was evaluated by using PEGMEMA monomers at $M_w \sim 500$. As shown in Fig. S1c,† the Young's modulus of the resulting PEGDA-co-PEGMEMA plastics decreased similarly when the PEGMEMA ($M_w \sim 500$) percentage was increased in the resin mixture, confirming the role of mono-functional PEGMEMA in the mechanical behavior of these plastics. The similarity between the Young's modulus of plastics derived from PEGMEMA at both ~300 and ~500 $M_w$ [*e.g.*, 9–10 times decrease in modulus when PEGMEMA content with either $M_w$ increased from 0% to 30%] suggests a dominant role of the mono-methacrylate groups over that of the $M_w$ on the elasticity.[46] The use of PEGMEMA ($M_w \sim 500$), however, yields more viscous PEGDA-co-PEGMEMA photopolymers due to the higher viscosity of this monomer compared to PEGMEMA at $M_w \sim 300$ (Fig. S3†). In SLA 3D-printing, higher resin viscosity leads to larger suction force applied onto the objects being printed,[47] which can compromise the fabrication of microfluidic devices with high-resolution or delicate overhangs (*e.g.*, the microchannel roof or a thin membrane). Thus, in the rest of the experiments, we used PEGMEMA at $M_w \sim 300$, simply referred to as PEGMEMA.

The design of microfluidic actuators requires a precise knowledge of the flexural behavior of the deflecting membrane materials; hence we characterized the flexural modulus of 3D-printed PEGDA-co-PEGMEMA. Fig. 1c illustrates a pressurization chamber used as the experimental setup to test the flexural deformation of 2 mm-diam., 60 μm-thick membranes from different polymer compositions. With this design, we 3D-printed three devices containing 0%, 20%, and 40% PEGDA-co-PEGMEMA membranes, and evaluated their transverse deflection under pneumatic actuation. As shown in Fig. 1d, under a constant pressure of 2 psi, the membrane deflection increased with increasing PEGMEMA content in the resin. This finding suggests that the addition of PEGMEMA to the resin decreases the flexural modulus of the printed membrane, which agrees well with a decreasing Young's modulus revealed in Fig. 1b. Next, we recorded membrane deflection over a wider pressure range (Fig. 1e). Within this pressure range, the deformation of the membranes follows the ideal elastic behavior of thin circular membranes[48] (fitted curves in Fig. 1e with $R_{avg}^2 \sim 0.98$).

Due to the viscoelastic nature of PEGDA-co-PEGMEMA plastics, hysteresis effects during continuous loading/unloading can lead to membrane failure by irreversible creep deformation or by fatigue rupture.[49] These behaviors are particularly enhanced when the externally induced membrane deflection is beyond its elastic elongation limit. Therefore, to mitigate this challenge, membrane deformation should be constrained by reducing the radius or by increasing the thickness of the membrane. Fig. S4† demonstrates the behavior of a 40% PEGDA-co-PEGMEMA membrane (800 μm diam. and 60 ± 10 μm thickness) with tolerant behavior towards applied constant and cycling loads without noticeable creep deformation after 12 h (Fig. S4c†) as well as sustained linear elastic behavior over $10^4$ cycles (Fig. S4d†) (note creep studies of polymer films for time scales of $t < 10^5$ s ≈ 12 h is considered a satisfactory short-term analysis; within this time frame many polymers applied in MEMS, *e.g.*, SU-8, have been fully characterized[50,51]).

Absorption of non-polar, hydrophobic molecules such as drugs and hormones into polymeric microfluidic chips limits their utility in biomedical applications such as drug screening, cell-based assays, and tissue engineering.[52] The uptake of such molecules reduces their effective concentration in the flowing reagents and may contaminate neighboring microchannels or subsequent perfusions.[53] PEGDA-co-PEGMEMA contains hydrophobic methacrylate groups from PEGMEMA, as revealed by the larger contact angle of 40% PEGDA-co-PEGMEMA surface compared to that of PEGDA (Fig. S5†). Since this hydrophobic surface could promote hydrophobic drug permeability, we measured the absorption behavior of this resin family. We 3D-printed a set of wells from PEGDA-co-PEGMEMA with 0%, 20%, and 40% PEGMEMA content in PEGDA (Fig. 2a), and incubated them with 1 mM solution of fluorescent Nile red as a representative hydrophobic molecule.[54] As revealed by the fluorescent intensity profiles in Fig. 2a, Nile red penetrates deeper into the walls as the PEGMEMA content in PEGDA-co-PEGMEMA increases, suggesting an increase in the molecular absorptivity of such resins. The observed behavior can likely be attributed to the enhanced hydrophobicity due to the increase in methacrylate groups with respect to PEGDA. To quantify the diffusion coefficient $D$ of Nile red in each well, the intensity profiles shown in Fig. 2b are fitted with Fick's law equation for diffusion in a semi-infinite wall, $c_{x,t} = \mathrm{erfc}\left(\frac{x}{\sqrt{4Dt}}\right)$ in which $c_{x,t}$ represents the normalized concentration of Nile red, $x$ is the distance from the edge, and $t$ denotes time. From these fittings, we obtained $D$ as $4.38 \times 10^{-8}$ cm$^2$ s$^{-1}$ (40% PEGDA-co-PEGMEMA), $1.28 \times 10^{-8}$ cm$^2$ s$^{-1}$ (20% PEGMEMA), and $0.64 \times 10^{-8}$ cm$^2$ s$^{-1}$ (0% PEGMEMA, *i.e.*, PEGDA). In other words, the diffusivity of Nile red increases with the percent of PEGMEMA in the 3D-printed part. Nevertheless, compared to PDMS where the diffusion coefficient of Nile red is $\sim 420 \times 10^{-8}$ cm$^2$ s$^{-1}$,[24] these $D$ values are much smaller (~100 times for 40% PEGDA-co-PEGMEMA and ~650 times for PEGDA, resulting in ~10 and ~26 times shorter diffusion lengths, respectively, than in PDMS), suggesting a great promise for using this resin family in 3D-printed drug-containing biomicrofluidic devices. Note that PEGDA-co-PEGMEMA is expected to have minimal non-specific adsorption of biomolecules due to the non-fouling properties of PEG.[55] To promote selective attachment of biomolecules, PEGDA-co-PEGMEMA resins can be mixed with various organic functional groups (*e.g.*, pyridyl disulfide[56]) to create thiol[56] or azide[57] conjugation anchors on the surface of the polymerized resin. With this approach, the 3D-printed device has a functionalized surface with high affinity towards biomolecules that bear a particular functional group (*e.g.*, thiol-bearing biomolecules in the case of pyridyl disulfide functionalized polymer surface).

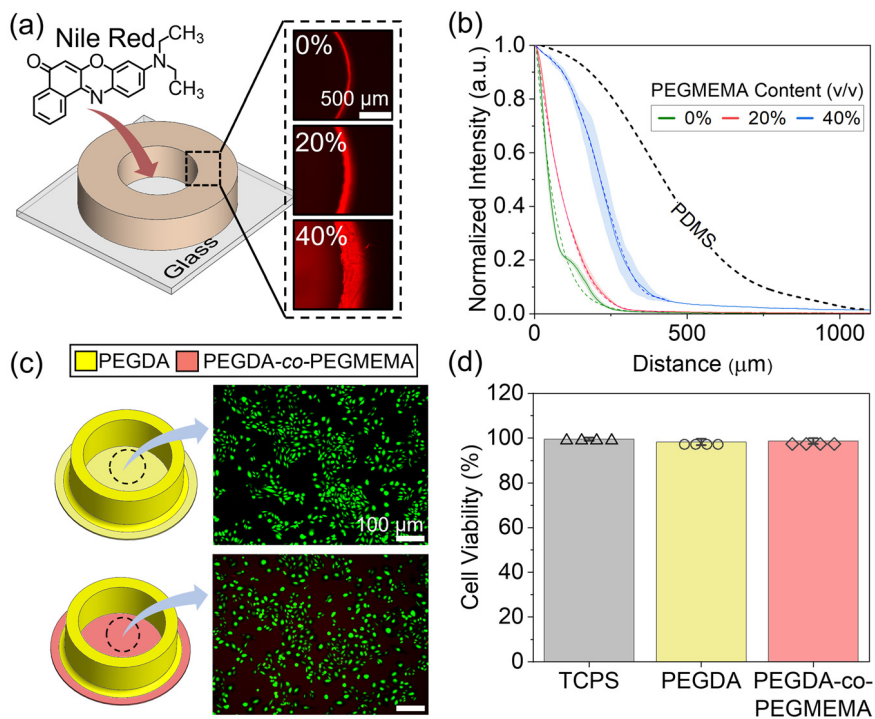

Fig. 2 Drug absorption and cytocompatibility of PEGDA-co-PEGMEMA plastics. (a) Illustration of the 3D-printed wells for absorptivity measurement along with fluorescent image of Nile red penetration into the wall. Here, 0%, 20%, and 40% refer to the content of PEGMEMA monomers in the PEGDA-co-PEGMEMA resin. (b) Fluorescent intensity profiles for the three images shown in (a) with respect to the distance from the edge of the well. The intensity profile for PDMS is from our previous work.[24] The dashed lines are curve fitting according to Fick's law equation for the semi-infinite wall. (c) Schematic of the two types of 3D-printed wells used in the cell culture study shown next to representative viability-stained fluorescent micrographs of CHO-K1 cultures, showing live (green) and dead (red) cells. The PEGDA-co-PEGMEMA resin contains 40% PEGMEMA. (d) Average cell viability counts obtained from micrographs ($n$ = 4 per condition) after 72 h of culture. The error bars represent standard error of the mean.

Cytocompatibility with acrylate-based photopolymers is a persistent challenge that hinders their applicability to SLA-printed microfluidic devices in cell-relevant studies.[58,59] The toxicity of the as-printed photopolymers mainly stems from the abundance of unreacted monomers on or near the surface.[22] The unpolymerized monomers require a stringent post-processing protocol to wash, cure, and sterilize the 3D-printed parts prior to bringing the part in contact with live cells or tissue. We evaluated the cytocompatibility of Chinese hamster ovary (CHO-K1) cells, a cell line often used in biotechnology and in biomedical research, with PEGDA-co-PEGMEMA printed plastics. For this purpose, we first 3D-printed wells with similar dimensions to a standard tissue culture polystyrene (TCPS) 24-well plate (~15 mm in diameter) (Fig. 2c). The bottom of the wells to which the cells attach is a 300 μm-thick layer printed in either PEGDA or 40% PEGDA-co-PEGMEMA. Following our previous work on PEGDA,[22] we then adopted a washing protocol (see Experimental section) to ensure the removal of cytotoxic unreacted monomers, photo-initiator, and photo-absorber compounds from the PEGDA and PEGDA-co-PEGMEMA plastics. Once the 3D-printed wells were treated, a known concentration of cells in DMEM media were seeded in each well (4 wells per condition) along with a control cell culture in TCPS. After 72 h, the viability of the cells was evaluated by a live/dead assay (Experimental section).

Fig. 2c shows a fluorescent micrograph of live (green), and dead (red) cells attached to PEGDA and PEGDA-co-PEGMEMA surfaces. There were virtually no dead cells on both PEGDA and 40% PEGDA-co-PEGMEMA wells (~97% viability, the same as in controls) after 72 h of culture (Fig. 2d). Although the biocompatibility of a SLA-printed part is ultimately dependent on the design, exact printing and post-printing protocols, the resin formulation details (e.g., the photo-absorber avobenzone has produced excellent cytocompatibility with PEGDA[19,28,60]), the type of tissue, and the duration of exposure to the tissue, our data suggests that the PEGDA-co-PEGMEMA resins combined with our post-printing washing protocol, results in prints of acceptable cytocompatibility for biomicrofluidic applications (where cells could possibly contact the walls made of resin, as the cells in Fig. 2 are contacting the surface made of resin).

While we have shown that PEGDA-co-PEGMEMA can sustain CHO cultures, the stability and biocompatibility of this material in any medical device would require further experimentation and validation, for ex. with a variety of cell types and conditions. PEGDA surfaces feature less protein adsorption[61] and drug absorption[24] than PDMS surfaces, but the long-term effects of cell culture or in vivo conditions on the mechanical properties of PEGDA-co-PEGMEMA remain to be studied.

Solvent compatibility was evaluated for printed polymers from PEGDA-*co*-PEGMEMA. First, we 3D-printed cubes of 3 mm in size using 0, 20, and 40% PEGDA-*co*-PEGMEMA. After doing post-processing on the printed cubes to remove uncured monomers (see section 4.2), the initial mass and volume of each cube were recorded (the volume was measured using microscope images and Fiji ImageJ analysis was used to estimate the cube dimensions). Then, the cubes were individually exposed to 1 mL of common laboratory solvents (Fig. 3). After 24 hours, the final volume of each cube was measured with the same method used to estimate the initial volume. To measure the final mass, the cubes were first dried in an oven at 50 °C for 12 hours, and their final masses were recorded afterwards.[35] Finally, the percentage increase in mass and in volume of each cube (3 cubes per resin type) was estimated using the initial and final mass/volume data. As seen in Fig. 3, all three PEGDA-*co*-PEGMEMA resins produce plastic parts with relatively small solvent uptake (less than 8% and 10% change in mass and volume, respectively) compared to 3D-printed PDMS, which can swell in volume up to 68% in THF and 13% in acetone.[35] More importantly, the swelling and mass increase in all PEGDA-*co*-PEGMEMA samples in DMSO (a widely used solubilizer for pharmaceutical drugs) is comparable to those exposed to water, suggesting an acceptable use of PEGDA-*co*-PEGMEMA for drug testing platforms.

## 2.2. 3P-printing

Fluidic actuators based on a deformable membrane enable reliable fluid control and transport in microfluidic systems,[1–4,25,62] allowing the large-scale integration of chemical and biological functions.[1,63] The performance of these actuators depends critically on the elastic behavior of the membrane that determines the actuation pressure and power conversion efficiency. Here we investigate the application of tunable PEGDA-*co*-PEGMEMA resins in designing and printing fluidic actuators with high performance. The tunability of the resin provides an extra degree of freedom in actuator design and optimization, as the desired maximum deflection of the dynamic membrane can be tailored not only by physical dimensions but also by modulating the ratio of PEGMEMA to PEGDA monomers according to Fig. 1b. To achieve an optimal structure, all the printed actuators in this work take advantage of a hybrid architecture that assembles a circular flexible membrane from PEGDA-*co*-PEGMEMA with a hard polymeric PEGDA housing. (Note that hybrid structures that combine a foreign membrane, *e.g.* from molded PDMS, with a 3D-printed structure[64] present additional challenges: (i) membrane insertion becomes more and more difficult as the desired objects decrease in size and increase in number; (ii) only flat geometries (membranes) can be easily inserted in practice; and (iii) PDMS does not bind to PEGDA, which compromises the device's seal.) To print such hybrid structures, we used 3P-printing,[45] a facile and reliable multimaterial printing method that uses a desktop SLA 3D-printer.[50] In 3P-printing (Fig. S6a†), the first step is to pause the 3D-printer during its operation at a desired moment when the resin is to be changed. During the pause, the surface of the semi-printed object is cleaned (twice) with an absorbing cellulose pad for ∼1 min to minimize contamination with the next resin. Due to the large wicking capacity of the cellulose pad, it can rapidly absorb the uncured resin while leaving unreacted monomers on the surface for bonding with the next material. Afterwards, the resin is switched to the next material (*e.g.*, by changing the build tray) followed by continuing with the rest of the print job or repeating these steps for adding a third resin. The effectiveness of the cleaning step in 3P-printing also depends on resin viscosity in the sense that a lower viscosity improves cleaning by enhancing the wicking flow rate into the absorbing pad.[65] To evaluate this step for PEGDA, we 3P-printed a simple two-resin part by sequentially printing PEGDA resins with and without fluorescent dyes. As shown by the fluorescent images in Fig. S6b and c,† this process led to minimal cross-contamination and a well-resolved interface. Owing to the comparable viscosity of PEGDA and PEGMEMA monomers (Fig. S3†), the cleaning

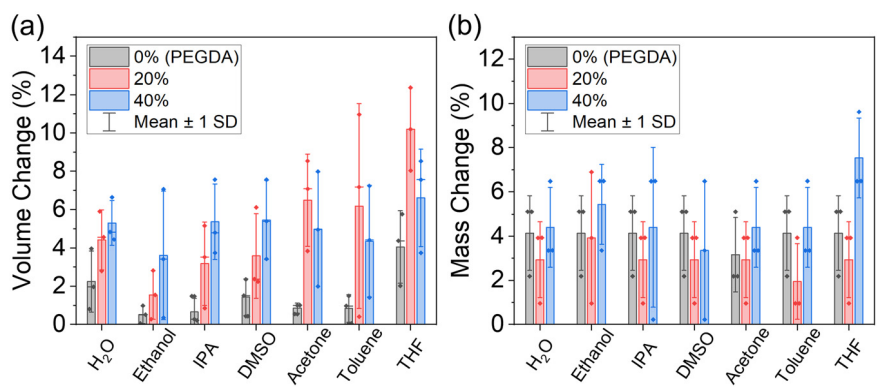

Fig. 3 Solvent compatibility of PEGDA-*co*-PEGMEMA plastics. (a) Volume and (b) mass increase in 3D-printed cubes (3 mm in size) from 0% (PEGDA), 20% and 40% PEGDA-*co*-PEGMEMA after exposure to common laboratory solvents for 24 hours. The error bars represent standard deviation.

procedure remains effective for all PEGDA-*co*-PEGMEMA resins.

The versatility of 3P-printing[45] allows for printing multiple acrylate resins side-by-side or on top of each other. As an example, we 3P-printed 0, 10, and 20% PEGDA-*co*-PEGMEMA resins in both planar (*x*–*y* plane) and layered (*z*-direction) arrangements using our SLA 3D-printer (Fig. 4a and b). Note that, although the interfaces between colored sections were thoroughly cleaned during 3P-printing, the abundance of unsaturated acrylate groups at these interfaces led to a strong adhesion between different polymers and prevented delamination. Therefore, most acrylate-based resins should be compatible with and bond well to PEGDA-*co*-PEGMEMA. Given the rapid curing rate of PEGDA-*co*-PEGMEMA family resins (characteristic curing time of ∼1 s, see Fig. S7†), the 3P-printing can be implemented to assemble sections from various PEGDA-*co*-PEGMEMA formulations into a large scale, monolithic architecture with sub-100 μm features. As an example, Fig. 4c illustrates a micrograph of a ∼15 mm-tall multimaterial Eiffel tower 3P-printed with 50 μm layer thickness and 75 μm *X*–*Y* resolution.

### 2.3. Multimaterial microfluidic actuators

Multimaterial 3P-printing can be used to improve the performance of SLA-printed microvalves by allowing the microfluidic designer to embed a flexible PEGDA-*co*-PEGMEMA membrane within a rigid PEGDA structure. In conventional soft lithographic devices pioneered by the Quake group, the membrane is made of a highly elastic PDMS layer (Young's modulus of ∼2 MPa (ref. 31 and 41)) interlayered between two orthogonal microchannels.[1] To simplify fabrication and allow 3D valve architecture, our group reported SLA-printed "Quake-style" microvalves made entirely in PEGDA plastic.[25] However, PEGDA is ∼400 times stiffer than PDMS, yielding microvalves that demand higher actuation pressure, larger membrane diameters and/or shallower seats for valve closure compared to conventional elastomeric devices. For any given printer, these shortcomings restrict the practical use and scalability of SLA-printed PEGDA microvalves due to geometrical and material design constraints. Here we employed our repertoire of PEGDA-*co*-PEGMEMA resins to 3P-print multimaterial valves with high performance and enhanced design freedom. The hybrid structure of these valves is comprised of a 1 mm-diam., 60 μm-thick flexible membrane from tunable PEGDA-*co*-PEGMEMA resins within a hard PEGDA plastic housing (Fig. 5a).

Fig. 5b shows a micrograph of a 3P-printed valve containing 40% PEGDA-*co*-PEGMEMA membrane along with the liquid and pneumatic channels. To determine the reliability of the valve, we applied inlet pressures between 0.1–0.4 psi and evaluated the valve closure by recording the outlet flow rate with a digital flow meter. Fig. 5c plots the output flow rate with respect to the control pneumatic pressure applied to the membrane. Using these curves, the actuation pressure of the valve is obtained as ∼3 psi for all inlet pressures below 0.4 psi in a 500 μm × 500 μm square-cross-section microchannel. To correlate this performance with membrane elasticity, we 3P-printed a set of geometrically identical devices with successively less flexible membranes made of 40%, 20%, and 0% PEGDA-*co*-PEGMEMA plastics. Fig. 5d compares the characteristic flow rate *versus* control pressure for these devices at 0.2 psi inlet pressure. These curves demonstrate that switching to a more elastic membrane significantly lowers the microvalve's actuation pressure, a key factor in determining the minimum size, thus increasing the scalability of the valve for a wider range of applications.[10] The actuation pressure of a microvalve primarily relies on the Young's modulus of the membrane which determines the maximum membrane deflection at a given actuation pressure (eqn (S1) in the ESI†).

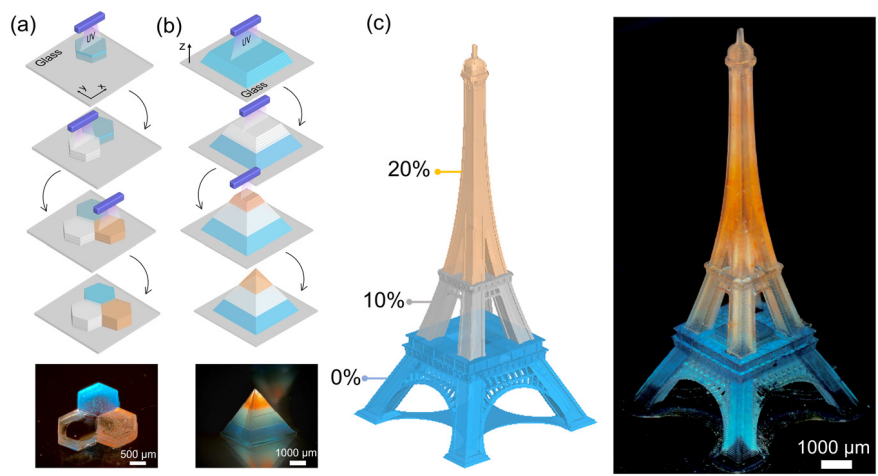

**Fig. 4** Multimaterial parts by 3P-printing. (a) Planar (*X*–*Y*) and (b) layer-by-layer (*Z*) arrangement of 0% (blue), 10% (white [transparent]), and 20% (orange) PEGDA-*co*-PEGMEMA resins. The resin is switched twice for every 100 μm printed layer to obtain the multimaterial hexagons, whereas it is only switched twice for printing the entire multimaterial pyramid. (c) High-resolution multimaterial Eiffel tower (∼15 mm-tall) fabricated by 3P-printing 0, 10, and 20% PEGDA-*co*-PEGMEMA resins.

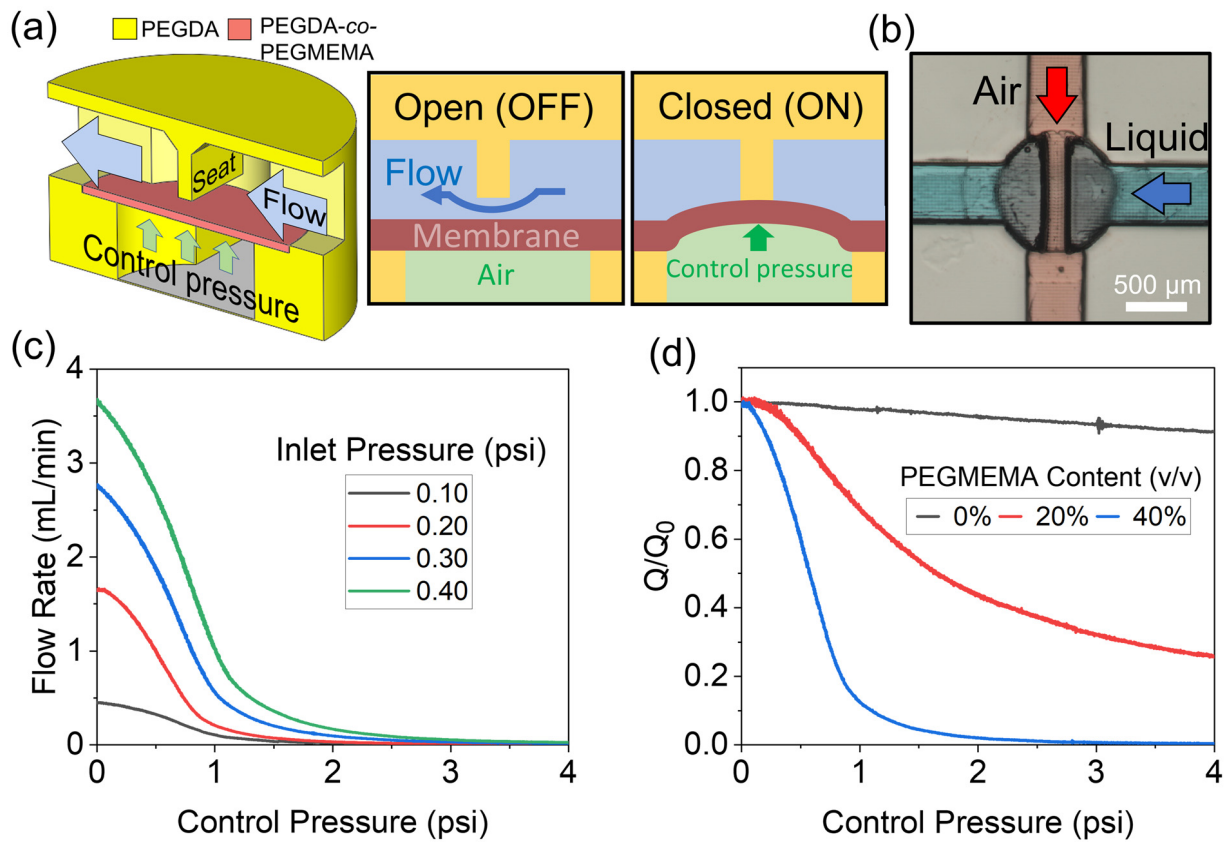

Fig. 5 PEGDA-*co*-PEGMEMA microvalves. (a) Cut-away view of the 3D model of the microvalve. The 2D schematics on the right show the valve closing mechanism (ON/OFF states). (b) Microscope image of the 3P-printed multimaterial valve. To visualize the pneumatic and liquid channels, here they are filled with red and blue dyes, respectively. Note that the valve is in an open state. (c) Flow rate *versus* control pressure for a microvalve 3P-printed with a 40% PEGDA-*co*-PEGMEMA membrane. (d) Normalized flow rate $Q/Q_0$ *versus* control pressure for three 3P-printed microvalves with membranes from 0%, 20%, and 40% PEGDA-*co*-PEGMEMA resins. Note that the thickness of the membranes in all microvalves is kept at 60 ± 10 μm, and the valve seat has a width of 100 μm.

Since 40% PEGDA-*co*-PEGMEMA and PDMS have comparable Young's modulus (Fig. 1b), membrane valves made from these two polymers have similar actuation pressures for any two given identical valve geometries.

The low viscosity of PEGDA-*co*-PEGMEMA is a key factor in allowing us to print high-resolution voids (*i.e.*, microchannels) and thin features (*i.e.*, overhanging membranes) needed for microfluidic actuators. For instance, the suction force during SLA process is proportional to the resin viscosity,[47] making it difficult to print stretchable membranes without causing significant initial slack. Moreover, unlike the more viscous elastomeric resins,[33–35] low-viscosity PEGDA-*co*-PEGMEMA can be easily washed away from the internal voids of the printed devices (*e.g.*, by applying a small pressure) during post-processing of the print, paving the way to print highly miniaturized fluidic actuators. As an example, Fig. S9† demonstrate our ability to build PEGDA-*co*-PEGMEMA into high-resolution microvalves with ~7-pixel (200 μm) microchannels, and ~15-pixel (400 μm) membrane size. As seen in Fig. S9d,† despite the scaled-down dimensions, the microvalves still function with low actuation pressure around ~6.5 psi.

Highly deformable PEGDA-*co*-PEGMEMA membranes can be exploited to displace large volumes of liquids for pumping applications. Thus, we fabricated peristaltic pumps by 3P-printing hybrid structures containing flexible PEGDA-*co*-PEGMEMA membranes and hard PEGDA housing. The micropumps (Fig. 6) consist of three circular chambers that are independently and cyclically actuated by three pneumatic lines, generating a net pumping mechanism akin to peristalsis. The first and third chambers are operated in a coordinated fashion to function as check-valves to rectify the flow. The role of the middle chamber is to displace the largest possible volume of liquid per stroke so it does not contain a valve seat unlike the first and last chambers (Fig. S10a†). A cross-section schematic of the micropump along with a microscopic image of the 3P-printed device is shown in Fig. 6a and b, respectively. To maximize the flow rate, the middle membrane is pressurized at a larger 4 psi of pressure compared to the adjacent check-valves with sequential opening/closing at 3 psi (Fig. S10b†).

The performance of the micropump, including the flow rate and its pressure delivery, relies on the flexibility of the three membranes; their flexibility determines the rate of volumetric displacement by the middle membrane and the reliable closure of the neighboring microvalves. To test this notion, we 3P-printed three micropumps containing 20, 30, and 40% PEGDA-*co*-PEGMEMA membranes, and evaluated their flow rate and maximum pumping pressure. A summary of the performance of the pumps is shown in Fig. 6c and d (see Experimental section for details of flow rate and pressure delivery measurements). Fig. 6c compares the output flow rates of the pumps with respect to the actuation phase interval $T_0$. The three graphs in Fig. 6c confirms that the inclusion of a more elastic membrane results in higher pump flow rates due to a larger volumetric displacement of liquid per actuation cycle. For instance, the pump with 40% PEGDA-*co*-PEGMEMA membranes can provide ~120% greater flow rate compared to 20% PEGDA-*co*-PEGMEMA membranes. To evaluate the pressure deliveries, we measured the maximum pressure of each device and compared them in Fig. 6d. Similar to the flow rate, changing the membrane material from 20 to 40% PEGDA-*co*-PEGMEMA leads to a large (~100%) increase in the maximum pumping pressure without the need to modify geometrical dimensions. The observed increase in the pump performance by a facile modulation of membrane material reveals higher design freedom for the 3P-printed pumps compared to their single-material counterparts.

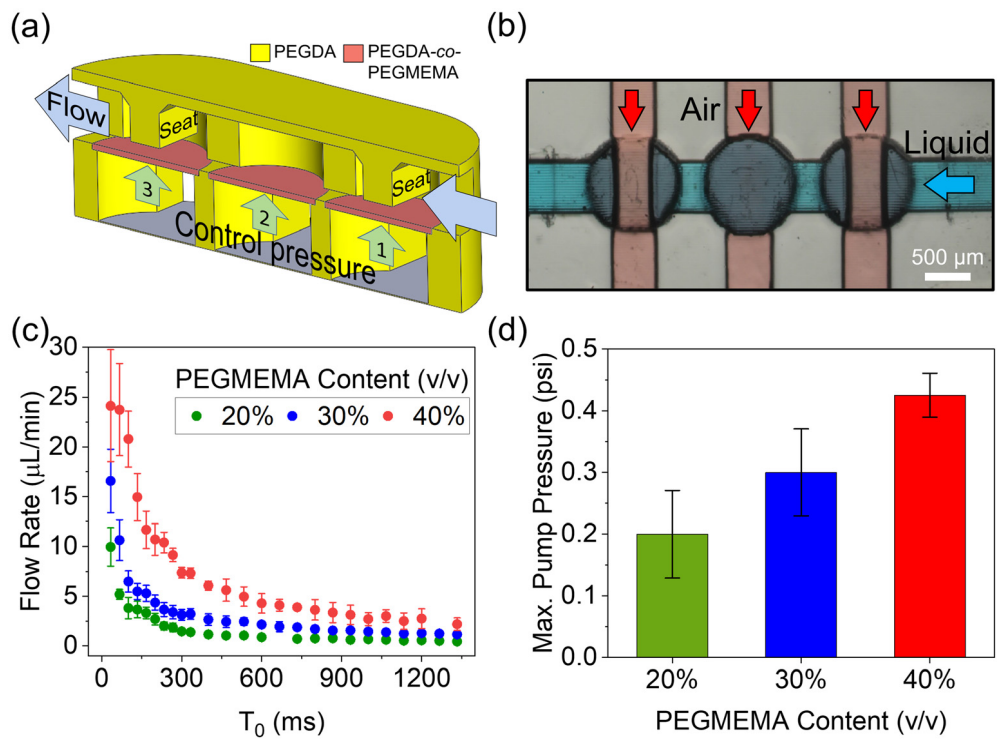

**Fig. 6** PEGDA-*co*-PEGMEMA peristaltic micropumps. (a) Cut-away view of the 3D model of the micropump. (b) Microscope image of the 3P-printed multimaterial peristaltic micropump. Note that the valves are in an open state. (c) Flow rate *versus* phase interval $T_0$ for 20, 30, and 40% PEGDA-*co*-PEGMEMA micropumps. (d) Maximum pressure delivery of the 3P-printed micropumps. Note that the thickness of the membrane in all micropumps is kept at 60 ± 10 μm. Moreover, the seats for the microvalves have a width of 300 μm. These wider seats provide a larger contact area for the membrane during rapid opening/closure.

Maintaining a constant flow rate in the presence of a fluctuating flow pressure is essential for many lab-on-a-chip devices in areas such as drug delivery, flow chemistry, and micro-dialysis, among others.[66] Flow regulators perform this task by adjusting the hydraulic resistance of a microfluidic system $R_h$ in harmony with the upstream flow pressure $\Delta p$ such that the flow rate $Q$ remains unchanged according to $\Delta p/Q = R_h$. To achieve this principle, active-mode regulators utilize externally applied pressure signal to increase (decrease) $R_h$ when $\Delta p$ increases (decreases). Here we demonstrate 3P-printed regulators that take advantage of a highly flexible PEGDA-*co*-PEGMEMA membrane to create a self-actuating effect without the need of an externally applied control. Fig. 7a illustrates a cross-sectional view of the 3D model of the regulator. In this hybrid structure, the regulator contains an embedded PEGDA-*co*-PEGMEMA membrane situated within a rigid PEGDA structure and a flow divider. During device operation, the flow that enters the regulator is split between the top and the bottom of the cylindrical chamber. Since the bottom section is closed, the quiescent liquid imposes a static pressure $P_b$ to the membrane's bottom. It can be simply shown that the median pressure difference between $P_b$ and the liquid pressure atop of the membrane $P_t$ ($P_b - P_t$) is ~$1/2\rho v^2$ in which $v$ denotes average flow velocity, and $\rho$ is the fluid density. This approximation implies that increasing (decreasing) the $v$ driven by the inlet pressure, passively lifts (relaxes) the membrane (Fig. 7b), leading to an increase (decrease) in the $R_h$ of the device. By a carefully chosen geometry, and the use of highly flexible 40% PEGDA-*co*-PEGMEMA membrane, we 3P-printed a multimaterial regulator (Fig. 7c) that exploits this phenomenon for flow stabilization. Fig. 7d plots the output flow rate of the regulator with respect to an increasing inlet pressure. Each curve corresponds to a device with a certain gap height $H$. As seen, the regulators with the 40% PEGDA-*co*-PEGMEMA membrane can maintain a constant flow rate (less than 6.5% variation) despite the increase in the inlet pressure from 0.8 to 2 psi. Meanwhile, the regulator with a stiff PEGDA membrane (dotted curve in Fig. 7d) is unable to perform this task, signifying the key role of flexible PEGDA-*co*-PEGMEMA plastics for successful flow regulation. Fig. 7d also reveals that the regulator's sensitivity can be tuned by adjusting the gap height $H$ such that a smaller $H$ decreases the stabilized output flow rate of the device, offering a straightforward approach to rapid-prototype a variety of microregulators for a wide range of applications.

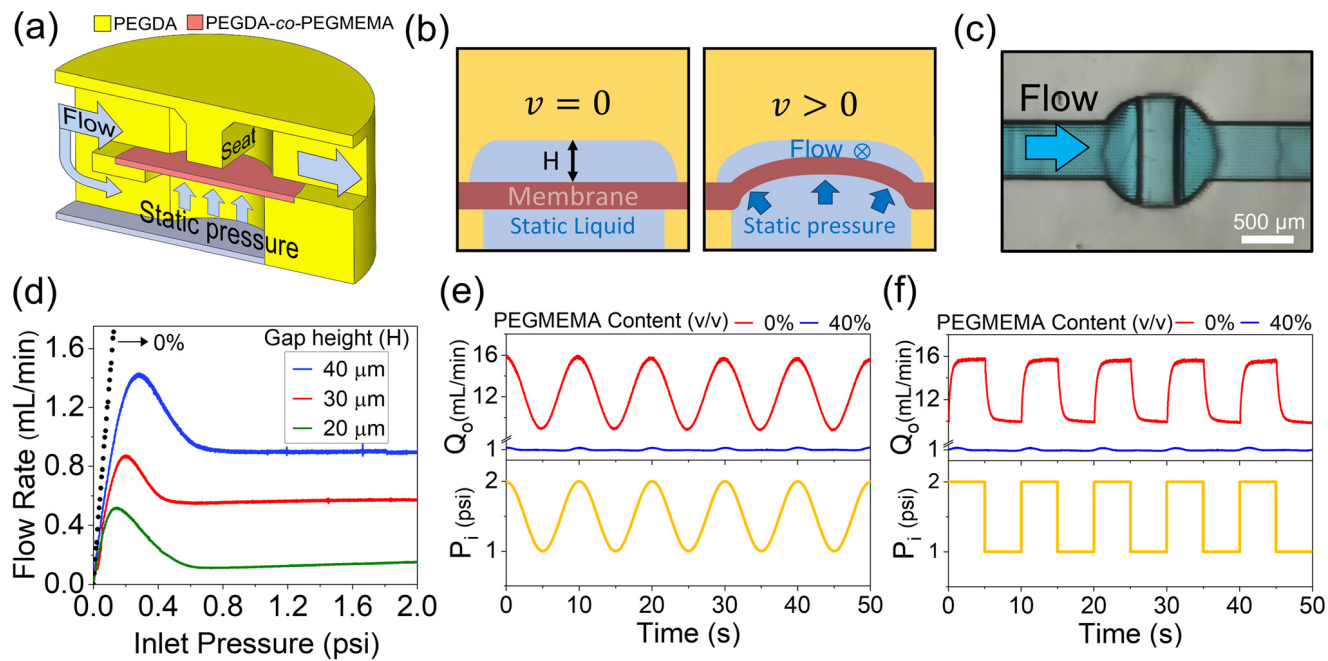

Fig. 7 PEGDA-*co*-PEGMEMA microregulators. (a) Cut-away view of the 3D model of the microregulator. (b) Cross-section schematics depicting the regulator self-actuating mechanism. (c) Microscope image of the 3P-printed multimaterial flow rate regulator with a 40% PEGDA-*co*-PEGMEMA membrane. (d) Flow rate *versus* inlet pressure for microregulators with different membrane gap height $H$. The three colored curves correspond to a regulator with a 40% PEGDA-*co*-PEGMEMA membrane. The dotted curve corresponds to the regulator with 0% PEGDA-*co*-PEGMEMA (*i.e.*, PEGDA) membrane with $H = 40$ μm. Regulator's output flow rate $Q_0$ in response to an oscillating input pressure $P_i$ in the forms of (e) sine wave and (f) step function. In these graphs, 0% refers to a device with rigid PEGDA membrane. Note that the thickness of the membrane in all regulators is kept at $60 \pm 10$ μm. Moreover, the seats have a width of 300 μm.

In many practical applications, the inlet pressure applied to a microfluidic system is of an oscillating nature.[67] Thus, we next evaluated the regulating performance of our 3P-printed device by applying wave-form pressure signals to the input flow. Fig. 7e and f plot the output flow rates of the regulator as a function of time (blue curves) alongside the response of a device made from rigid PEGDA (*i.e.*, 0% PEGDA-*co*-PEGMEMA) (red curves). The pressure signals applied to the inlets are also shown on the bottom graphs (yellow curves). As shown, both sine wave and step-function signals were investigated to evaluate the performance of the device under distinct wave formats. Under both input pressure functions, the regulator with a 40% PEGDA-*co*-PEGMEMA membrane maintains a constant flow rate despite the large 1 psi fluctuation in the input pressure. The output flow rate varies within only ∼7% and ∼8% of the average flow rate for the case of a sinusoidal and a step-function pressure input, respectively (Fig. 7e and f). On the other hand, the device with a rigid PEGDA membrane (red curves) is unable to stabilize the flow rate, and the output flow rates follow the wave forms of the input pressure signal.

Compared to existing regulators based on PDMS microfabrication,[66,68] the 3P-printed regulator significantly reduces fabrication time and cost by bypassing layer bonding and alignment steps. Additionally, our regulator design takes advantage of the 3D space to shrink the device's footprint. For instance, the pressurization chamber of the regulator is accommodated under and alongside the main flow channel. This architecture allows the 3P-printed regulator to occupy the same space as a microvalve.

## 3. Conclusions

Microfluidic actuators enable automated manipulation of small fluid samples at high throughput for chemical and biological applications. This class of microfluidic devices have had a huge impact in biomedical research and biotechnology, yet their soft lithographic fabrication is complex, and their dissemination presents multiple challenges. Until now, however, microfluidic actuators have been produced exclusively by microfluidic engineers. The advent of high-resolution SLA 3D-printing offers the possibility of democratizing the fabrication of microfluidic actuators (and biomicrofluidic devices in general), lowering the cost and barriers to access of this key technology. Despite the promise of 3D-printing, present SLA-printed microfluidic actuators are limited to a single polymeric resin, restricting the researcher's ability for device design and optimization. While FDM and MJM technologies have been used to build multimaterial microfluidic actuators,[69–71] both have limitations for biomicrofluidic device fabrication. FDM suffers from low resolution, z-direction anisotropy, and layer-by-layer appearance.[72] In MJM, the inks are often limited to non-cytocompatible inks,[23] non-Newtonian shear-thinning UV-curable polymers, and post-processing is needed to ensure the removal of solid sacrificial material from the 3D-printed microchannels.[73]

Here we have combined a family of acrylate-based photopolymer, PEGDA-co-PEGMEMA, with 3P-printing, a high-resolution multimaterial 3D-printing technique compatible with desktop SLA printers, to fabricate all-plastic fluidic actuators. The performance of these actuators was shown to be superior to any single-polymer actuator of the same size because the membrane, printed in tunable PEGDA-co-PEGMEMA resins, is a highly flexible plastic compared to conventional PEGDA.[25,28] (Note that our demonstration of PEGDA-co-PEGMEMA microactuators are compared to PEGDA alone.) For instance, we showed a drastic reduction in actuation pressure of a 3P-printed multimaterial fluidic valve, facilitating scalability of this component without compromising its geometry. The PEGDA-co-PEGMEMA resin family allows facile adjustment of the Young's modulus of the printed plastics (by ~400-fold, reaching that of PDMS) through a change in the constituent monomer ratio. This capability unlocks further freedom in design and optimization of fluidic actuators by a straightforward tuning of PEGDA-co-PEGMEMA resins. Importantly, while the elastic behavior of PEGDA-co-PEGMEMA can be easily modulated, other desirable physical and biochemical properties (i.e., low viscosity, solvent compatibility, low drug permeability compared to PDMS, and cytocompatibility) remained suitable for printing high-resolution (bio)microfluidic devices. We envision that the combination of 3P-printing with our tunable resins will facilitate the integration of high-performance, practical multimaterial actuators in 3D-printed fluidic chips by microfluidic engineers elsewhere.

## 4. Experimental

### 4.1. Preparation of photopolymers

PEGDA photopolymers are prepared by mixing PEGDA monomer (Sigma-Aldrich) with 0.6 wt% phenylbis (2,4,6-trimethylbenzoyl) phosphine oxide (Irgacure-819, BASF, IL) as the photo-initiator and 0.6 wt% 2-isopropylthioxanthone (ITX) (PL Industries, PA) as the photo-absorber compound. We prepared PEGDA-co-PEGMEMA resins by mixing 0.6 wt% of Irgacure-819 and 0.6 wt% of ITX in a monomer precursor containing a varying volumetric ratio of PEGDA and PEGMEMA at $M_w$ ~ 300 (Sigma-Aldrich). The colored resins are prepared by adding ~10 mg mL$^{-1}$ of food-coloring dyes to the photopolymers, followed by a vigorous vortex mixing step.

### 4.2. 3D-printing process

We used a desktop DLP-based SLA 3D-printer (Pico 2 HD, Asiga) with an X–Y pixel resolution of 27 μm, equipped with a 385 nm UV-LED light source that has a projection light intensity of 85 mW cm$^{-2}$. To 3D-print microfluidic devices, we used a glass slide attached to the 3D-printer's build platform. Prior to 3D-printing, the glass slide is washed successively with acetone, IPA, and DI-water, and dried in an oven at 80 °C for 20 min. The glass slide is finally silanized overnight to ensure the attachment of photopolymerized resins onto its surface, according to our previously published protocol.[25] The glass is attached to the build platform using a drop of PEGDA resin followed by curing under ambient light for ~5 min. To detach the glass slide from the build platform when the print is complete, we wedged a razor blade between the build platform and the glass slide.[25] To post-process microfluidic devices, first we immersed the 3D-printed device (usually attached to the glass slide) in a water bath covered away from the ambient light for ~20 min. Afterwards, we flushed the channels with soapy water followed by IPA and DI-water to ensure the removal of residual uncured resins from the microchannels. After the parts are dried in ambient air, we then placed them under a UV light source for one minute to completely cure the resin. This step improves the structural integrity of the parts. The STL files for all 3P-printed devices are available in the ESI.† Note that the PPP process is a manual step, and depending on the individual it takes 1–2 min to swap the resin, clean the interface, and switch to a new resin. All the devices in this study are printed with 50 μm layer thickness, and the overall printing time is ~1 hour.

### 4.3. Measurement of tensile properties

To measure the Young's modulus of the printed polymers, we 3D-printed dog-bone specimens (Fig. S1a†) and recorded their stress–strain curves (Fig. S1b and c†) using Instron 5584H Load Frame equipped with a 2 kN load cell by following the standard ASTM protocol for evaluating plastics. During the test, the longitudinal displacement rate was set at 10 mm min$^{-1}$, and the elongation was recorded with an extensometer until the sample breaks. To be an acceptable measurement, the breakage line must lie between the two shoulders within the narrowest section of the specimen (Fig. S1a†). Unfortunately, the dog-bone samples 3D-printed from 40% PEGDA-co-PEGMEMA failed because the metal clamps of the testing apparatus applied too much pressure on the material, leading to unacceptable break lines. Therefore, the Young's modulus of this plastic was estimated according to the thin membrane theory proposed in ref. 48 (see Fig. S2†). Note that, at 50% PEGMEMA and beyond, the photo-cured polymers are very soft and lose structural integrity, corresponding to the abundance of monoacrylated PEGMEMA in the resin. Thus, we chose 40% to be the practical limit of PEGMEMA addition to the PEGDA resin. To estimate the elasticity of the samples with 0 to 35% PEGDA-co-PEGMEMA, the linear portion of the stress–strain curves for each plastic was fitted with a straight line. The slope of this line was reported as the Young's modulus of the material.

### 4.4. Viscosity measurement of PEGDA and PEGMEMA

Viscosities of PEGDA and PEGMEMA ($M_w$ ∼ 300 and 500) monomers were measured using a rotational rheometer (Anton Paar MCR 301) equipped with a 25 mm flat disk. The tests were conducted at 25 °C, and the viscosities were recorded for shear rates below 1000 s$^{-1}$. In this range, PEGDA and PEGMEMA ($M_w$ ∼ 300 and 500) demonstrate Newtonian behavior (Fig. S3a†). Thus, the average of the recorded viscosities in Fig. S3b† was reported as the viscosity for each monomer.

### 4.5. Cytocompatibility test

To evaluate the cytocompatibility of the PEGDA and PEGDA-co-PEGMEMA plastics, we used Chinese hamster ovary cells (CHO-K1) cultured in DMEM media (Invitrogen) supplemented with 10% fetal bovine serum, 1% penicillin–streptomycin (Hyclone) and 2 mM L-glutamine (Sigma-Aldrich). In the first step, we 3D-printed a set of wells (4 wells per material) with PEGDA and 40% PEGDA-co-PEGMEMA photopolymers (Fig. 2). Next, to ensure the removal of toxic monomers or residual resin components, these wells were washed with IPA for 2 h, cured under UV light in a water bath for 12 h, and finally treated with oxygen plasma (75 mTorr, 30 s, 10 W) to improve cell attachment.[22] After this treatment protocol, 10$^4$ cells (in 500 μL of DMEM media) were seeded in each 3D-printed well, alongside a 24 well plate as the control experiment. The wells were then incubated under 5% $CO_2$ at 37 °C for 72 h. On the third day, the cells were washed with PBS, and the DMEM medium was replaced by Live Cell Imaging Solution (Invitrogen) containing 2 μM calcein green and 1 μM ethidium homodimer (Invitrogen) for fluorescent staining of live/dead cells.

### 4.6. Characterization of 3P-printed actuators

To evaluate the 3P-printed microvalve, the pneumatic channel was connected to a computer-controlled pressure source (Elveflow OB1 MK3), and the output flow rate is recorded with a flow meter (Sensirion AG). To operate the micropump, each pneumatic channel is connected to a separate channel on the pressure source. To measure the micropump's flow rate, we recorded video with a light microscope at each phase interval $T_0$ and used it to calculate the amount of time it takes for the pump to displace liquid from the first marker to the last one engraved onto the 3P-printed device (ESI† video). To calculate the maximum pump pressure, we operated each pump at its highest flow rate (lowest $T_0$) while applying a back pressure with an external pressure source. The back pressure at which the meniscus of the liquid stops moving is recorded as the maximum pressure delivered by the micropump.[25] During devices' operation, we observed no bubble intrusion into the flow channel. This is likely due to the low gas permeability of PEGDA-co-PEGMEMA membranes compared to PDMS (see ref. 68 and 74). Measurements of gas permeability are underway.

## Conflicts of interest

The authors declare no competing financial interest.

## Acknowledgements

The authors acknowledge the financial support from the National Institute of General Medical Sciences (1R21GM137161) and the National Cancer Institute (2R01CA181445). The authors also thank Dr. Lisa F. Horowitz for her guidance on cell culture experiments.